# Gravitational Lensing and the Mass Distribution of Clusters

**Matthias Bartelmann**[1,2] **and Ramesh Narayan**[1]

[1] Harvard-Smithsonian Center for Astrophysics, 60 Garden Street, Cambridge, MA 02138, USA; [2] Max-Planck-Institut für Astrophysik, Karl-Schwarzschild-Strasse 1, D–85740 Garching, FRG

8 November 1994

## 1 Introduction

The possibility that clusters of galaxies may act as efficient gravitational lenses was discussed by Noonan (1971) and Dyer & Roeder (1976) even before the first gravitational lens was discovered (Walsh, Carswell & Weymann 1979). Paczyński & Gorski (1981) interpreted a group of three high-redshift QSOs as multiple images of a single QSO split by two intervening galaxy clusters, and Narayan, Blandford & Nityananda (1984) suggested that cluster lensing may be as important as lensing by galaxies if clusters have small core radii. Webster (1985) first noticed that extended background objects may be coherently distorted by foreground galaxy clusters, and that this effect could be used to constrain cluster mass distributions. Hammer & Nottale (1986) argued that the magnification by galaxy clusters could measurably affect cosmological information obtained from the Hubble diagram of brightest cluster galaxies. Blandford, Phinney & Narayan (1987) simulated lensing by a cluster-scale mass and showed that arcs and arclets will result.

Despite these early theoretical realizations that lensing by clusters of galaxies may be common and could provide important information, the subject moved to the forefront only after the first giant blue luminous arcs were discovered in the clusters A 370 and Cl 2244−02 (Lynds & Petrosian 1986, 1989, Soucail et al. 1987a,b). The interpretation of these arcs in terms of gravitational lensing (Paczyński 1987) was confirmed when the first arc redshifts were measured (Soucail et al. 1988, Miller & Goodrich 1988, Mellier et al. 1991). The further development of techniques to detect smaller arclets (Tyson, Valdes & Wenk 1990) has now created a very fertile and promising field of research which is providing valuable information on the mass distribution of clusters.

The chief advantage of gravitational lenses in the study of clusters is that the lensing effect depends directly on the total mass distribution, whereas most other methods need to make assumptions about how the luminous matter is distributed relative to the mass, and about the dynamical state it is in. The giant arcs provide information about the total masses (and hence the mass-to-light ratios) of the lensing clusters, and impose constraints on the symmetry of the cluster potentials, their core radii, radial density profiles, and the frequency of substructure. Using arclets, methods have been developed recently which even allow us to reconstruct full two-dimensional mass maps of the lensing clusters. We summarize these results in the following sections.

For general information on gravitational lensing see the review by Blandford & Narayan (1992) or the monograph by Schneider, Ehlers & Falco (1992). A detailed review on arcs and arclets in galaxy clusters has been recently compiled by Fort & Mellier (1994).

## 2 Basic Gravitational Lens Optics

Following the thin-lens approximation, which is valid in basically all situations of astrophysical interest, the mass distribution of the lens can be projected onto a lens plane orthogonal to the line-of-sight from the observer to the source. Another plane, the source plane, is constructed parallel to the lens plane at the distance of the source. It is convenient to trace light rays backwards on straight lines from the observer to the lens plane, where they are deflected, and to continue them backwards on straight lines from the lens plane to the source plane.

If the surface mass density $\Sigma$ of the lens somewhere becomes comparable to or exceeds the critical value, $\Sigma_{\rm crit} \approx 0.35$ g cm$^{-2}/D$(Gpc), then caustics will be formed on the source plane (e.g., Blandford & Narayan 1986). Here, $D = D_{\rm d}D_{\rm ds}/D_{\rm s}$ where $D_{\rm d,s,ds}$ are angular-diameter distances from the observer to the lens, observer to the source, and from the lens to the source, respectively. Sources inside caustics are split into multiple images. If an extended source like a galaxy happens to touch a caustic, then an arc is formed (Grossman & Narayan 1988, Blandford & Kovner 1988, Narayan & Grossman 1989, Kovner 1989). A galaxy touching a fold caustic produces a moderate arc, while one which overlaps a cusp or the related beak-to-beak or lips caustic, produces a giant arc. If the lens is subcritical, or if the source lies outside the caustics, then the image of the source is only weakly distorted and is called an arclet. Illustrations of these different kinds of imaging situations are shown in the following figure.

To sufficient accuracy, the deflection of light rays by a lens is described by a two-dimensional potential,

$$\psi = \frac{2D_{\rm ds}}{D_{\rm d}D_{\rm s}} \int \Phi \, dz \,, \tag{1}$$

where $\Phi$ is the three-dimensional Newtonian potential of the lens and the integration is performed along the line-of-sight. The local properties of the lens mapping are contained in the curvature matrix or Hessian $(\psi_{,ij})$, where the comma preceding an index denotes partial differentiation with respect to the corresponding coordinate. The trace-free antisymmetric part of $(\psi_{,ij})$ measures the distortion of images and is called the shear matrix, with independent components $\gamma_1 = \gamma \cos 2\varphi$ and $\gamma_2 = \gamma \sin 2\varphi$, where $\varphi$ is the orientation of the shear. The symmetric part measures the isotropic magnification or demagnification of images and is called the convergence $\kappa$. Thus

$$(\psi_{,ij}) = \begin{pmatrix} \kappa + \gamma \cos 2\varphi & \gamma \sin 2\varphi \\ \gamma \sin 2\varphi & \kappa - \gamma \cos 2\varphi \end{pmatrix} \,. \tag{2}$$

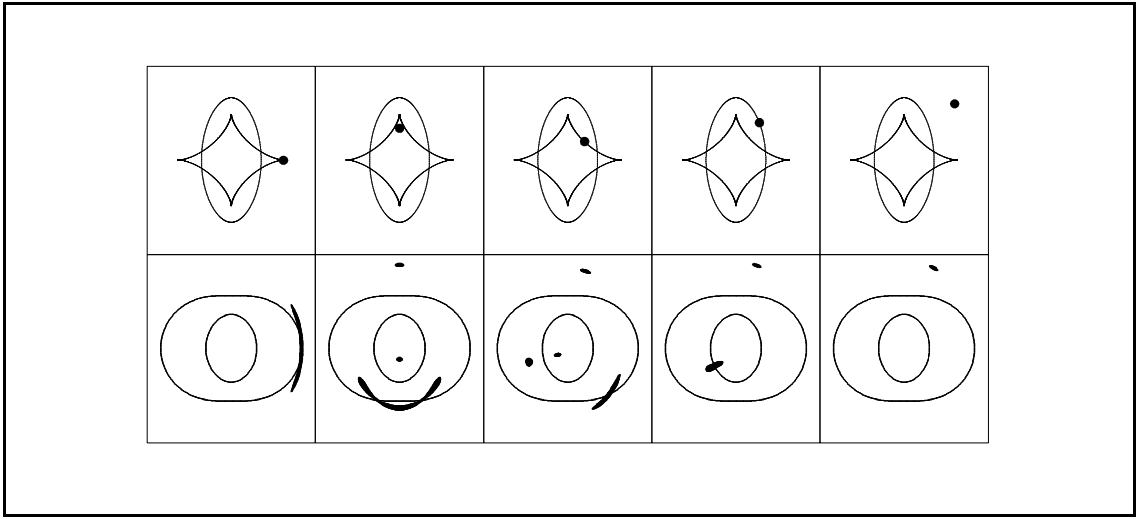

**Fig. 1.** Illustration of five different imaging situations for an elliptical lens with finite core. The top row shows the position of the source relative to the caustics, the bottom row shows the images of the source relative to the critcal curves. From left to right, the columns display: source on a "naked cusp" forming a large arc without counterarc; source inside a cusp forming a large arc from three merging images and counterarcs; source on a fold forming a shorter arc from two merging images; source on the radial caustic forming a short radial arc; source outside the caustic forming an arclet. The sizes of the outer and the inner critical curves approximately correspond to the Einstein radius and the core radius of the cluster, respectively. The shapes of large arcs shown in the first three panels are characteristic of elliptical lenses. Spherically symmetric lenses behave very differently (e.g., Grossman & Narayan 1988).

Note that, by Poisson's equation, $\kappa = (\psi_{,11} + \psi_{,22})/2 = \Sigma/\Sigma_{\rm crit}$, where $\Sigma$ is the local surface density of the lens. The axis ratio of an arc(let) is given by

$$\frac{\text{semi-minor axis}}{\text{semi-major axis}} = \frac{1 - \kappa - \gamma}{1 - \kappa + \gamma} , \qquad (3)$$

so that by measuring image distortions we can obtain information on $\gamma$.

## 3 Enclosed Mass and Mass-to-Light Ratio

If a large arc is found in a galaxy cluster, the total cluster mass inside a circle with radius equal to the distance $\theta_{\rm arc}$ of the arc from the cluster center can be roughly estimated to be

$$M \approx 1.1 \times 10^{14} \, M_\odot \left(\frac{\theta_{\rm arc}}{30''}\right)^2 \left(\frac{D}{1 \text{ Gpc}}\right) . \qquad (4)$$

This result follows from assuming that $\theta_{\rm arc}$ measures the angular Einstein radius of the cluster (cf. Fig. 1). Equation (4) can be converted to an estimate for the line-of-sight velocity dispersion $\sigma_v$ of the cluster,

$$\sigma_v \approx 10^3 \left(\frac{\theta_{\rm arc}}{28''}\right)^{1/2} \left(\frac{D_{\rm s}}{D_{\rm ds}}\right)^{1/2} \text{ km s}^{-1} . \qquad (5)$$

Of course, to use these formulae, we need to know the redshifts of both the lens and the source. Eqautions (4) and (5) provide only rough estimates of $M$ and $\sigma_v$. More reliable values are obtained with detailed lens models. Table 1 lists representative results for masses, velocity dispersions, and mass-to-blue-light ratios $M/L_B$ of three arc clusters, obtained by means of detailed lens models. More results of this kind can be found in Fort & Mellier (1994). The accuracy of these mass estimates is hard to evaluate, but they are probably correct to a factor of two.

**Table 1.** Examples of masses, mass-to-blue-light ratios, and velocity dispersions derived from lens models for three arc clusters.

| Cluster | $M$ ($M_\odot$) | $M/L_B$ (solar) | $\sigma_v$ (km/s) | Reference |
|---|---|---|---|---|
| A 370 | $\approx 10^{14}$ | $\approx 200$ | $\approx 1350$ | Grossman & Narayan (1989)<br>Bergmann et al. (1990)<br>Kneib et al. (1993) |
| A 2390 | $\approx 1.5 \times 10^{14}$ | $\approx 120$ | $\approx 1250$ | Pello et al. (1991) |
| MS 2137 | $\approx 5 \times 10^{13}$ | $\approx 250$ | $\approx 1100$ | Mellier et al. (1993) |

## 4 Asphericity of Clusters

The fact that the observed giant arcs never have counter-arcs of comparable brightness, and rarely have even small counter-arcs, implies that the lensing geometry has to be non-spherical (e.g., Grossman & Narayan 1988, Kovner 1989; cf. Fig. 1). Therefore, cluster potentials must have substantial quadrupole and perhaps also higher multipole moments. In the case of A 370, for example, there are two cD galaxies, and the potential quadrupole estimated from their separation is consistent with the quadrupole required to model the observed giant arc (Grossman & Narayan 1989). The more detailed model of A 370 by Kneib et al. (1993) shows a remarkable agreement between the lensing potential and the strongly aspheric X-ray emission of the cluster.

Large deviations of the lensing potentials from spherical symmetry also help increase the probability for producing large arcs. Bergmann & Petrosian (1993) argued that the apparent abundance of large arcs relative to small arcs and arclets can be reconciled with theoretical expectations if aspheric lens models are taken into account. Bartelmann, Steinmetz & Weiss (1994) showed that the probability for large arcs can be increased by more than an order of magnitude if aspheric cluster models with significant substructure are used instead of smooth spherically symmetric models.

## 5 Core Radii

If a cluster is able to produce large arcs, its surface-mass density in the core must be approximately supercritical, $\Sigma \gtrsim \Sigma_{\mathrm{crit}}$. If applied to simple lens models, this condition requires

$$\theta_{\mathrm{core}} \lesssim 15'' \left(\frac{\sigma_v}{10^3 \text{ km/s}}\right)^2 \left(\frac{D_{\mathrm{ds}}}{D_{\mathrm{s}}}\right).$$

Narayan et al. (1984) were the first to show that cluster mass distributions need to have smaller core radii than those derived from optical and X-ray observations if they are to produce strong effects of gravitational lensing. This message has been confirmed many times by later efforts to model giant arcs. In virtually every case the core radii estimated from lensing are significantly smaller than the estimates from optical and X-ray data. Some representative results on lens-derived core radii are listed in Table 2, where the estimates correspond to $H_0 = 50 \text{ km s}^{-1}\text{Mpc}^{-1}$.

**Table 2.** Limits on cluster core radii derived from modelling large arcs.

| Cluster | $r_{\mathrm{core}}$ (kpc) | Reference |
|---|---|---|
| A 370 | $\lesssim 60$ | Grossman & Narayan (1989) |
|  | $\lesssim 100$ | Kneib et al. (1993) |
| MS 2137−23 | $\approx 50$ | Mellier et al. (1993) |
| Cl 0024+1654 | $\lesssim 130$ | Bonnet et al. (1994) |
| MS 0440+0204 | $\ll 90$ | Luppino et al. (1993) |

Statistical analyses based on spherically symmetric cluster models lead to similar conclusions. Miralda-Escudé (1992, 1993a) argued that cluster core radii can hardly be larger than the curvature radii of large arcs. Wu & Hammer (1993) claimed that clusters either had to have singular cores or density profiles much steeper than isothermal in order to reproduce the abundance of large arcs. Although this conclusion can substantially be weakened once deviations from spherical symmetry are taken into account (Bartelmann et al. 1994b), it remains true that $r_{\mathrm{core}} \lesssim 100$ kpc is required in all observed clusters, and cores of this size can also be reconciled with large-arc statistics.

Interestingly, there is at least one observation which indicates that cluster cores, although small, must be finite. Fort et al. (1992) discovered a radial arc near the center of MS 2137−23. To produce a radial arc, the core radius has to be roughly equal to the distance between the cluster center and the radial arc (cf. Fig. 1). Mellier, Fort & Kneib (1993) find $r_{\mathrm{core}} \gtrsim 40$ kpc in MS 2137−23. Bergmann & Petrosian (1993) presented a statistical argument in favor of finite cores by showing that lens models with singular cores produce fewer large arcs (relative to small arcs) than observed. The relative abundance increases with a small finite core.

# 6 Radial Density Profile

Many of the observed giant arcs are unresolved in the radial direction, some of them even with excellent seeing. Since the faint blue background galaxies, which provide the source population for the arcs, can be resolved (e.g., Tyson, these proceedings), the giant arcs appear to be demagnified in width. It was realized by Hammer & Rigaut (1989) that spherically symmetric lenses can radially demagnify giant arcs only if their radial density profiles are steeper than isothermal; the maximum demagnification is obtained for a point mass lens, where it is a factor of two. Kovner (1989) and Hammer (1991) demonstrated that, irrespective of the mass profile and the symmetry of the lens, the demagnification is given by $\approx 2(1-\kappa)$, where $\kappa$ is the convergence at the position of the arc. Thin arcs therefore require $\kappa \lesssim 0.5$. Since giant arcs have to be located close to those critical curves in the lens plane along which $1 - \kappa - \gamma = 0$, large thin arcs additionally require $\gamma \gtrsim 0.5$.

In principle, the radius of curvature of large arcs relative to their distance from the cluster center can be used to constrain the steepness of the radial density profile (Miralda-Escudé 1992, 1993a), but results obtained from observed arcs are not yet conclusive (Grossman & Saha 1994). One problem with this method is that substructure in clusters tends to enlarge curvature radii irrespective of the mass profile of the dominant component of the cluster (Miralda-Escudé 1993a, Bartelmann et al. 1994b).

Wu & Hammer (1993) argued for steep mass profiles on statistical grounds, because the observed abundance of large arcs appears to require highly centrally condensed cluster mass profiles in order to increase the central mass density of clusters while keeping their total mass constant. However, their conclusions are based on spherically symmetric lens models and are significantly altered when the symmetry assumption is dropped (Bartelmann et al. 1994b).

It should be kept in mind that not all arcs are thin. Some "thick" arcs are known (e.g., in A 2218, Pello-Descayre et al. 1988, and in A 2390, Pello et al. 1991), and it is quite possible that thin arcs predominate just because they are more easily detected than thick ones due to observational selection effects. Also, Miralda-Escudé (1992, 1993a) has argued that intrinsic source ellipticity can increase the probability of producing thin arcs, while Bartelmann et al. (1994b) showed that the condition $\kappa \lesssim 0.5$ which is required for thin arcs can be more frequently fulfilled in clusters with substructure where the shear is larger than in spherically symmetric clusters.

The measurement of a coherent weak shear pattern out to a distance of $\lesssim 1.5$ Mpc from the center of the cluster Cl 0024+1654 by Bonnet, Mellier & Fort (1994) demonstrates a promising method of constraining cluster mass profiles. Those observations show that the density decreases rapidly outward, but the data are compatible both with the isothermal profile and a steeper de Vaucouleurs profile.

# 7 Mass Sub-condensations

A 370 has two cD galaxies and is a clear example of a cluster with multiple mass centers. A two-component mass model centered on the cD galaxies (Kneib et al. 1993) fits the lens data as well as X-ray and deep optical images of the cluster. A 2390 is an interesting example because it contains a straight arc (Pelló et al. 1991, see also Mathez et al. 1992) which can be produced only with either a lips or a beak-to-beak caustic (Kassiola, Kovner & Blandford 1992). If the arc is modeled with a lips caustic, it requires the mass peak to be close to the location of the arc, but this is not where the cluster light is maximum. With a beak-to-beak caustic, the model requires two separate mass condensations, one of which could be at the peak of the luminosity, but then the other has to be a dark condensation.

A 370 and A 2390 are the most obvious examples of what is probably a widespread phenomenon, namely that clusters are in general not fully relaxed but have substructure as a result of ongoing evolution. If clusters are frequently substructured, this can lead to systematic effects in the statistics of arcs and in the derived cluster parameters (Bartelmann et al. 1994b, Bartelmann 1994b).

## 8 Arclets and Mapping the Cluster Surface Density

Giant arcs are relatively rare, but almost all clusters with velocity dispersions $\sigma_v \gtrsim 500$ km/s produce measurable distortions of background galaxy images (Webster 1985, Miralda-Escudé 1991a). Since at faint magnitudes ($B \approx 27$) there are as many as $\approx 100$ faint galaxies per square arc minute (Tyson 1988), most of which probably are at fairly large redshifts $z \approx 0.7 - 1.4$ (Guhathakurta, Tyson & Majewski 1990, Fort 1992, Colless et al. 1993), this means that it is possible to obtain a huge amount of information from image distortions induced by foreground clusters. Indeed, deep images of cluster fields reveal large numbers of arclets (e.g., Tyson et al. 1990, Smail et al. 1991) and one of the most exciting areas today involves the use of these arclets to derive information about the foreground mass.

The idea of using arclet data to study the mass distributions of clusters was suggested first by Webster (1985) and again by Grossman & Narayan (1989) after the discovery of the faint blue galaxies by Tyson (1988). Tyson et al. (1990), Kochanek (1990) and Miralda-Escudé (1991a) showed how it is possible to determine the cluster center, to constrain parametrized cluster models, and to map the cluster surface density, and Tyson et al. (1990) actually obtained some preliminary results on a few clusters. Later, Kaiser & Squires (1993) described a well-defined algorithm for constructing surface-density maps of clusters from coherent distortions of galaxy images in the field. This method, frequently called "cluster lens inversion", has recently been discussed and refined in great detail (Schneider & Seitz 1994, Seitz & Schneider 1994, Kaiser 1994; see also the articles by Tyson and Squires in this volume).

The reconstruction technique proposed by Kaiser & Squires is based on the fact that the convergence $\kappa$ and the shear components $\gamma_1, \gamma_2$ of a lens are combinations of second derivatives of the same scalar potential $\psi$ [cf. Eq. (2)]. Therefore they are not independent but are related through relations such as

$$\kappa(\boldsymbol{\theta}) = -\frac{1}{\pi} \int d^2\theta' \left[ \mathcal{D}_1(\boldsymbol{\theta} - \boldsymbol{\theta}')\gamma_1(\boldsymbol{\theta}') + \mathcal{D}_2(\boldsymbol{\theta} - \boldsymbol{\theta}')\gamma_2(\boldsymbol{\theta}') \right] , \qquad (7)$$

where the convolution kernels $\mathcal{D}_1, \mathcal{D}_2$ are given by

$$\mathcal{D}_1(\boldsymbol{\theta}) = \frac{\theta_1^2 - \theta_2^2}{|\boldsymbol{\theta}|^4} \;,\; \mathcal{D}_2(\boldsymbol{\theta}) = \frac{2\theta_1\theta_2}{|\boldsymbol{\theta}|^4} \;. \tag{8}$$

Kaiser & Squires (1993) realized that in the weak-shear limit, $|\gamma| \ll 1$, the shear can be directly obtained from the distortion of galaxy images. The intrinsic ellipticity of galaxies, however, introduces noise into the reconstruction, requiring appropriate smoothing of the data. The method appears feasible in principle, as demonstrated by Fahlman et al. (1994), Smail et al. (1994), and Bonnet et al. (1994). It suffers, however, from intrinsic degeneracies and systematic effects. Schneider & Seitz (1994) showed that, if the shear is not weak, the coherent ellipticities of galaxy images measure a combination of $\kappa$ and $\gamma$ rather than $\gamma$ alone, and Kaiser (1994) and Seitz & Schneider (1994) extended the Kaiser & Squires method into the strong-lensing regime. Since the area integral in Eq. (7) formally extends over the complete plane, the finiteness of the area covered by arclet observations introduces boundary effects which are addressed by Schneider (1994). Even after these systematic effects are taken into account, the distortion pattern caused by a lens is still invariant under the transformation

$$\kappa \to (1 - \lambda) + \lambda\kappa$$

with arbitrary $\lambda \neq 0$, because adding a sheet of constant surface-mass density to the lens while simultaneously rescaling $\kappa$ and $\gamma$ does not change the distortion pattern (Schneider & Seitz 1994). This degeneracy can be removed if $\kappa$ can be determined at least at one point in the cluster field by another method, e.g., if large arcs are present or if a change in the number counts of galaxies behind the cluster can be observed (Broadhurst, Taylor & Peacock 1994).

The reconstruction method by Kaiser & Squires has been employed to estimate the mass distributions of a few clusters. For the cluster MS 1224, the estimated mass is higher than the virial mass by a factor $\approx 2 - 3$ (Fahlman et al. 1994, Carlberg et al. 1994), while for the cluster A 2163, the hottest and most luminous X-ray cluster in the sky, the reconstruction is surprisingly compatible with there being no mass at all in the cluster (Squires, these proceedings). The method requires careful investigation before it can reliably be used to determine absolute cluster masses rather than relative cluster mass maps.

# 9 Comparison of Lensing Results with other Determinations

## 9.1 Enclosed Mass

Three different methods are currently used to estimate cluster masses. Galaxy velocity dispersions yield a mass estimate from the virial theorem. The X-ray emission of rich galaxy clusters is dominated by thermal free-free emission and therefore depends on the squared density of the intracluster gas, which in turn traces the gravitational potential of the clusters. Such estimates usually assume that the cluster gas is in thermal hydrostatic equilibrium and that the potential is at least approximately spherically symmetric. Finally, large arcs in clusters provide a mass estimate through Eq. (4) or by more detailed modeling.

These three mass estimates are in qualitative agreement with each other up to factors of $\approx 2-3$.

Miralda-Escudé & Babul (1994) compared X-ray and large-arc mass estimates for the clusters A 1689, A 2163 and A 2218. They took into account that these clusters deviate from spherical symmetry, so that their masses may deviate from the estimate provided by Eq. (4), and obtained lensing masses from individual lens models which reproduce the observed arcs. They arrived at the conclusion that in A 1689 and A 2218 the mass required for producing the large arcs is higher by a factor of $2-2.5$ than the mass required for the X-ray emission, and proposed a variety of reasons for such a discrepancy, among them projection effects or non-thermal pressure support. Loeb & Mao (1994) specifically suggested that strong turbulence and magnetic fields in the intracluster gas may constitute a significant non-thermal pressure component in A 2218 and thus render the X-ray mass estimate too low.

Bartelmann (1994b) showed that cluster mass estimates obtained from large arcs by straightforward application of Eq. (4) are systematically too high by a factor of $\approx 1.6$ on average, and by as much as a factor of $\approx 2$ in 1 out of 5 cases. This discrepancy arises because Eq. (4) assumes a smooth spherically symmetric mass distribution whereas realistic clusters are asymmetric and have substructure. Note that Daines, Jones & Forman (1994) found evidence for at least two separate mass condensations along the line of sight towards A 1689, while the arclets in A 2218 show at least two mass concentrations. It appears that cluster mass estimates from lensing require detailed lens models in order to be accurate to better than $\approx 30-50\%$. In the case of MS 1224, Fahlman et al. (1994) and Carlberg et al. (1994) have obtained masses using the Kaiser & Squires weak-lensing cluster reconstruction method. Their mass estimates are $2-3$ times higher than the cluster's virial mass. Carlberg et al. find evidence from velocity measurements that there is a second poor cluster in the foreground of MS 1224 which may explain the result. All of these mass discrepancies illustrate that cluster masses must still be considered uncertain to a factor of $\approx 2$.

## 9.2 Core Radius

Lensing estimates of cluster core radii are generally much smaller than the core radii obtained from optical or X-ray data. Note, however, that the X-ray core radii depend on whether the cooling regions of clusters are included into emissivity profile fits or not, because the cooling radii are of the same order of magnitude as the core radii. If they are included, the best-fit core radii are reduced by a factor of $\approx 4$ (Gerbal et al. 1992, Durret et al. 1994). The estimates from lensing are fairly robust and probably reliable. The results generally suggest that the dark matter is much more centrally condensed than the luminous matter. Many clusters with large arcs have cD galaxies, so the cD may be suspected to be the cause of the small core. However, there are also non-cD clusters with giant arcs, e.g., A 1689 and Cl 1409 (Tyson 1990) and MS 0440+02 (Luppino et al. 1993). In fact, Tyson (1990) claims that the former two clusters have cores smaller than 100 kpc, just like the other arc clusters containing cD galaxies.

### 9.3 Does Mass Follow Light?

Leaving the core radius aside, does the mass follow the light? It is clear that the mass cannot be as concentrated within the galaxies as the optical light is (e.g., Bergmann, Petrosian & Lynds 1990). However, if the optical light is smoothed and assumed to trace the mass, then the resulting mass distribution is probably not very different from the true mass distribution. For instance, in A 370, the elongation of the mass distribution required for the giant arc is along the line connecting the two cD galaxies in the cluster (Grossman & Narayan 1989) and in fact Kneib et al. (1993) are able to achieve an excellent fit of the giant arc and several arclets with two mass concentrations surrounding the two cDs. Their model potential also agrees very well with the X-ray emission of the cluster. In MS 2137, the optical halo is elongated in the direction indicated by the arcs for the overall mass asymmetry (Mellier et al. 1993), and in Cl 0024, the mass distribution is elongated in the same direction as the galaxy distribution (Wallington, Kochanek & Koo 1994). Smail et al. (1994) find that the mass maps of two clusters reconstructed from weak lensing agree fairly well with their X-ray emission. An important counterexample is the cluster A 2390, where the straight arc requires a completely dark mass concentration (Kassiola et al. 1992).

### 9.4 What Kinds of Clusters Produce Giant Arcs?

Which parameters determine whether or not a galaxy cluster is able to produce large arcs? Clearly, large velocity dispersions and small core radii favor the formation of arcs. As argued before, intrinsic asymmetries and substructures also increase the ability of clusters to produce arcs because they increase the shear and the number of cusps.

The abundance of arcs in X-ray luminous clusters appears to be higher than in optically selected clusters, and this might suggest that X-ray brightness of the cluster is the critical factor that determines how efficient a cluster will be in lensing. At least a quarter, maybe half, of the 38 X-ray bright clusters selected by Le Fèvre et al. (1994) contain large arcs, while Smail et al. (1991) find only one large arc in a sample of 19 distant optically selected clusters. However, some clusters which are prominent lenses (A 370, A 1689, A 2218) are moderate X-ray sources, while other clusters which are very luminous X-ray sources (A 2163, Cl 1455) are poor lenses. This shows that, while there probably is a correlation between X-ray brightness and the ability to produce strong lensing, the correlation is not at all tight especially if very prominent arcs are considered. In fact, substructure appears to be at least as important as X-ray brightness. For example, A 370, A 1689 and A 2218 all seem to have clumpy mass distributions.

Another possibility is that giant arcs preferentially form in clusters with cD galaxies. A 370, for instance, even has two cDs. However, non-cD clusters with giant arcs are known, e.g., A 1689, Cl 1409 (Tyson et al. 1990), and MS 0440+02 (Luppino et al. 1993).

## 10 Weak Lensing by Large-Scale Structures in the Universe

Lensing by even larger scale structures than clusters has been discussed. Fugmann (1990) found that there is an excess of Lick galaxies in the vicinity of high-redshift, radio-loud QSOs, and showed that the excess extends to $\approx 10'$ around the QSOs. If the excess is real, it is most likely caused by the magnification bias due to gravitational lensing, and the scale of the lens must be very large; galaxy-sized lenses have Einstein radii of a few arc seconds and are therefore irrelevant for this kind of association.

Bartelmann & Schneider (1993a) showed that current models of large-scale structure formation can explain large-scale QSO-galaxy associations, provided a double magnification bias (Borgeest, von Linde & Refsdal 1991) is assumed, and they confirmed correlations of high-redshift, radio-loud QSOs with optical and infrared galaxies and with the diffuse X-ray emission (Bartelmann & Schneider 1993b, 1994; Bartelmann, Schneider & Hasinger 1994). The result by Bartelmann & Schneider (1993b) was recently confirmed by Seitz & Schneider (1994b), who investigated correlations between the same sample of QSOs and Zwicky clusters, following in part an earlier study by Rodrigues-Williams & Hogan (1994). Bartelmann (1994a) has shown that constraints on the density perturbation spectrum and the bias factor of galaxy formation can be obtained from the angular cross-correlation function between QSOs and galaxies.

The idea of looking for the weak lensing action of large-scale mass distributions through the distortions induced on background galaxy images was first suggested by Kristian & Sachs (1966) and Gunn (1967) and was analyzed in more detail by Babul & Lee (1991), Jaroszyński et al. (1990), Miralda-Escudé (1991b), Blandford et al. (1991), Bartelmann & Schneider (1991) and Kaiser (1992). The effect is expected to be weak (magnification and shear are typically on the order of $5 - 10\%$), and a huge number of galaxies has to be imaged with utmost care before a coherent shear signal can be observed. Mould et al. (1994) have set a limit on the observed distortion in a blank field. Their limit is consistent with standard models of structure formation. Bonnet et al. (1993) found weak shear in the vicinity of the twin quasar 2345+007 and confirmed the presence of a dark cluster-like mass concentration. This is a promising technique to check for the presence of suspected dark matter clumps.

# 11 Concluding Comments

Gravitational lensing has proved to be a powerful technique to measure the amount and distribution of mass in clusters of galaxies. Using lensing, it is possible to estimate the total gravitating mass, the core size, the density profile, the mass distribution, and the amount of substructure in clusters. With the advent of methods to utilize data on arclets, lensing promises to develop into an even more powerful diagnostic for cluster studies. Comparative studies of clusters using their lens effect, their X-ray emission, and measurements of the dynamics of the cluster constituents promise to yield unambiguous information about the structure and the evolution of clusters.